\newcommand{\beq} {\begin{eqnarray}}
\newcommand{\eeq} {\end{eqnarray}}
\documentclass[12pt, preprint,preprintnumbers,amsfonts,aps,nofootinbib]{revtex4}

\begin{document}

\title{Reheating Closed String Inflation}

\author{Daniel Green}
\email{drgreen@stanford.edu}
\affiliation{SLAC and Department of Physics, Stanford University, Stanford, CA 94305-4060} 

\preprint{SU-ITP-07/10}
\preprint{SLAC-PUB-12693}

\begin{abstract}
Protecting the inflationary potential from quantum corrections typically requires symmetries that constrain the form of couplings of the inflaton to other sectors.  We will explore how these restrictions affect reheating in models with UV completions.  In particular, we look at how reheating occurs when inflation is governed by closed strings, using N-flation as an example.  We find that coupling the inflaton preferentially to the Standard Model is difficult, and hidden sectors are typically reheated.  Observational constraints are only met by a fraction of the models.  In some working models, relativistic relics in the hidden sector provide dark matter candidates with masses that range from keV to PeV, with lighter masses being preferred.
\end{abstract}
\maketitle
\section{Introduction}
A crucial part of any model of inflation is the successful reheating of the Standard Model.  Despite the obvious importance of this process, there is typically little concern that it will not occur.  From the bottom-up standpoint, there are good reasons for this view.  The first is that it is trivial to add couplings that allow the inflaton to decay, in toy models.  The second is that the scale of inflation is usually very high compared to the scale of BBN.  For this reason, the reheat temperature is rarely (if ever) at risk of being too low for BBN.  Finally, unlike the phase of inflation that precedes it, the physics of reheating leaves little observational signature.  There are hopes for some signatures of reheating to appear in gravitational waves \cite{Khlebnikov:1997di,Easther:2006vd,Felder:2006cc, Dufaux:2007pt,GarciaBellido:2007dg,GarciaBellido:2007af}, but not in any currently measurable range. However, from the existence of problems like gravitino overproduction \cite{Weinberg:1982zq,Kawasaki:1994af,Khlopov:1984pf,Kallosh:1999jj,Giudice:1999yt}, it is already clear that the top-down view towards reheating is less favorable.

There is good reason to look at these problems from the top-down in inflationary models.  Inflation can be sensitive to UV physics, given that high-scale physics can ruin the slow-roll of inflation.  Protecting the slow-roll parameters from high-scale effects can dramatically change one's perspective on what is generic or likely.  For example, top-down considerations have pointed to possible large non-gaussianities \cite{Alishahiha:2004eh, Silverstein:2003hf} and observable cosmic superstrings \cite{Sarangi:2002yt,Copeland:2003bj,Dvali:2003zj}.  In general, one may hope for other low-scale observables that contain hints of UV physics.  Open string inflation \cite{Baumann:2007ah,Kachru:2003sx} is one example in which reheating could potentially offer such clues.  In this model, reheating occurs through tachyon condensation, so relic gravitons\cite{Kofman:2005yz,Frey:2005jk,Chialva:2005zy,Cline:2002it} or KK modes\cite{Chen:2006ni} might be observable.  However, one might think this is a special case, due to its effectively string scale reheat temperature.

There is one key difference between bottom-up and top-down approaches: coupling to hidden sectors.  A common feature in SUSY extensions of the Standard Model and UV completions via string theory is the appearance of hidden sectors that couple very weakly to the Standard Model.  In SUSY model building, these are typically needed for SUSY breaking.  However, the inflaton is typically assumed\footnote{Decay to the Standard Model can be ensured by having the inflaton charged under the Standard Model gauge group \cite{Allahverdi:2006iq,Allahverdi:2006we}.  These types of models will fall into the category of `open string inflation' from the type II description used in this paper.} to decay only to the Standard Model.  If these additional sectors are not sufficiently unstable\footnote{In models like \cite{Kitano:2006wz,Ibe:2006rc}, reheating the SUSY breaking sector restores supersymmetry \cite{Craig:2006kx}, thus making it problematic regardless of the stability of particles.}, then reheating them could be a serious problem (e.g., \cite{Kolb:1985bf,Kim:1991mv,Berezhiani:1995am, Berezhiani:2000gw, Dine:2006ii,Endo:2007ih,Endo:2006xg}, although much of this work is done in specific models that do not contain a realistic inflationary sector).

In this paper, we study reheating in closed string models using N-flation as an example.  We will find that the inflaton generically couples to all sectors, and that one must tune couplings in order to avoid bounds from BBN.  In hidden sectors with hot\footnote{Here `hot' means relativistic at freeze-out, not relativistic at BBN or during structure formation} stable relics (which includes all theories with mass gaps below the reheat temperature), the relic is typically a dark matter candidate.  However, avoiding over-closure requires substantial tuning if the relic mass is much greater than 1 MeV.

We begin with a discussion of reheating in chaotic inflation, focusing on how reheating changes when we try to account for new physics at the Planck scale.  This will lead us to consider N-flation as a UV completion.  In section III, we specifically consider the reheating of N-flation \cite{Dimopoulos:2005ac}.  We will measure the tuning required to meet experimental bounds depending on the types of relics and the overall energy density in the hidden sector.  In section IV, we will discuss the features of N-flation that generalize to other closed string models.  

\section{Chaotic Inflation}
We would like to understand the role of a UV completion in inflation and reheating.  With this in mind, we will look at how reheating depends on the steps taken to protect the potential against quantum corrections from UV physics.  Let us look at chaotic inflation \cite{Linde:1983gd} as an example, returning to generalizations later.  At the level of the classical Lagrangian, chaotic inflation looks like the simplest possible model for inflation.  The frequent starting point for such models is the Lagrangian
\beq
\label{classical}
\mathcal{L} =\frac{1}{2}(\partial_{\mu}\phi \partial^{\mu}\phi-m_{\phi}^2 \phi^2)-g^2 \phi^2 \chi^2-h\phi\bar{\psi}\psi+\ldots
\eeq
where $m_{\phi} \approx 10^{-6} M_{pl}$ by using ${\delta \rho \over \rho} \approx 10^{-5}$ from the CMB.  The inflaton is coupled to a scalar $\chi$ and a fermion $\psi$, both with masses in the range $100 GeV \ll m \ll m_{\phi}$.  These particles are required to be unstable and to decay only to Standard Model particles.  All Lagrangians we write down are understood to be minimally coupled to gravity.

With no higher-order terms, one can achieve many more e-folds than the necessary 60; $\langle \phi \rangle$ is typically bounded by the requirement that the energy density does not exceed $M_{pl}$.  Reheating proceeds through both a phase of broad parametric resonance to the $\chi$ fields, know as preheating \cite{Traschen:1990sw,Kofman:1994rk} (if $g \geq 10^{-6}$), and a phase of direct decay to $\psi$ particles.  The preheating phase is capable of producing gravitational waves that might be visible to future experiment \cite{Khlebnikov:1997di,Easther:2006vd,Felder:2006cc}.

The first requirement that is often imposed is radiative stability.  A model will be defined as stable if quantum corrections (including the renormalization of existing couplings) do not dramatically alter the behavior of the model.  Here, we use a UV cutoff at the Planck scale to regulate divergences.  This cutoff could be lower, which would alter a few formulas.  However, we will see that the Planck scale is the important scale in the models we are considering.  If we consider one-loop diagrams with $\chi$ running in the loop, all even powers of $\phi$ are generated.  The most stringent requirement comes from the mass term for $\phi$ itself.  This one-loop diagram is quadratically divergent, giving a contribution to the mass of order $\frac{g^2}{16 \pi^2} M_{pl}^2$.  Consistency with the CMB requires $m_{\phi}\simeq 10^{-6} M_{pl}$ and so implies that $g < 10^{-5}$.  This requirement makes the period of preheating ineffective\footnote{In the literature, the mass is often fine tuned to $10^{-6} M_{pl}$, and then one requires that the potential is sufficiently flat, giving a constraint of $g < 10^{-3}$}.  A similar argument requires that $h < 10^{-5}$.  The reheating of this model is now predominantly through the elementary decay\footnote{For $10^{-6}<g<10^{-5}$ there is a window where broad resonance still operates.  However, in this window, there is also a mass on the order of $m_{\phi}$ generated for $\chi$, which impedes particle production.} of $\phi$ and narrow resonances in $\chi$.

Because gravity is non-renormalizable, we expect to see new physics appear at the scale $M_{pl}$.  Therefore, we should treat (\ref{classical}) as an effective field theory below that scale and include all allowed higher-dimension operators, suppressed by appropriate powers of $M_{pl}$, with order one coefficients.  These are not contributions from gravity below the scale (which do not contribute order one coefficients) but rather due to particles (or something else) of mass greater than $M_{pl}$ that have been integrated out.  These contributions give a potential for $\phi$ of the form
\beq
V(\phi) = \sum_{n} \lambda_{n} M_{pl}^{4-2n} \phi^{2n}.
\eeq
Since chaotic inflation requires $\phi > M_{pl}$, if any $\lambda$ is order one then slow roll will not proceed.  I will describe theories as natural when the small sizes of couplings are due to symmetries (exact or approximate).

In order to get a natural theory of chaotic inflation, one can use shift symmetries to control the higher-dimension operators.  One such model, known appropriately as `Natural Inflation' \cite{Freese:1990rb}, uses an axion with $f> M_{pl}$ to ensure that a flat potential is obtained. In particular, the action is constrained to be of the form
\beq
\mathcal{L} = \frac{1}{2}\partial_{\mu}\phi \partial^{\mu}\phi + \Lambda^4 \cos({\phi \over f})+{\phi \over f} F\wedge F+\ldots,
\eeq
where $\Lambda$ is some dynamically generated scale (which will be fixed to get the right spectrum) and $F$ is some gauge field which will be taken to be part of the standard model.  This model inflates much like any theory of chaotic inflation does.  However, now we are required to reheat through a dimension five operator, and the resulting reheat temperature is given by
\beq
T_{RH} \propto {\sqrt{M_{pl} m_{\phi}^{3}} \over {f}}.
\eeq
For $f>M_{pl}$ one will get $T_{RH} < 10^9$ GeV. 

Thus, for models with similar physics during inflation, we see significantly different reheating depending on how the potential is protected from quantum corrections.  However, most natural models are protected by global symmetries, which are expected to be broken in any UV completion of gravity.  For this reason, having a technically natural model does not necessarily prevent the appearance of dangerous operators suppressed by powers of $M_{pl}$.  Therefore, it is relevant to ask if a given natural model can be embedded in such a completion.  For the purposes of this paper, we will take this completion to be string theory, as it is currently the only theory with answers to these questions.

In order to produce Natural inflation in string theory, one would need to find axions with $f> M_{pl}$.  However, this seems impossible \cite{Svrcek:2006yi,Banks:2003sx}.  The best one might be able to do is use $N$ axions to produce a natural model of assisted inflation \cite{Liddle:1998jc,Mazumdar:2001mm,Copeland:1999cs,Jokinen:2004bp} known as N-flation \cite{Dimopoulos:2005ac}.  Ideally, one would like to have a large number of axions with $f \approx M_{pl}$ and equal masses on the order of $m_{\phi}$, which would collectively do the job of a single inflaton moving a super-planckian distance in field space.  Degenerate masses could be hard to achieve, but a generic compactification may be close enough.  In particular, it was shown in \cite{Easther:2005zr} that, given the form of the axion mass matrix in some compactifications with stabilized moduli, the eigenvalue distribution for a generic compactification can be modeled by random matrices.  The resulting distribution of masses turns out to be sufficient for successful inflation, given appropriate initial conditions.

\section{Reheating N-flation}
In the previous section we introduced N-flation, a closed string model of assisted inflation.  Here we will expand on the string construction, as described in \cite{Easther:2005zr}, for the purpose of discussing the physics of reheating in this model.  In type IIB models there are $h_{1,1}$ axions associated with the four-cycles $B_{4}^{i}$ via
\beq
\phi_i = \int_{B_{4}^{i}} C_4 .
\eeq
In a supersymmetric compatification, these axions combine with the volumes of the four cycles, $\rho_i$, to form the complex parameter $\tau_i \equiv \rho_i-i\phi_i$.  A non-perturbative superpotential is generated from either Euclidean D3 branes wrapped on the four cycle or from gaugino condensation on D7 branes wrapping the four cycles giving a superpotential of the form
\beq
W= W_{0} + \sum_{i} A_i e^{-a_i \tau_{i}}.
\eeq
Here $W_{0}$ is the flux-induced superpotential, and $a_{i}$ is either $2 \pi$ for Euclidean D3s or $\frac{2 \pi}{M}$ for D7 brane gaugino condensation.  These axions have the usual coupling to four-dimensional gauge fields, through the Chern-Simons terms for the gauge fields that live on D7 branes that wrap the four cycles.  Both the kinetic terms and the mass terms for the axions are non-diagonal in the $\tau_{i}$ basis.  Once diagonalized, this theory behaves like N independent axions with a mass distribution that follows the Marcenko-Pastur Law.

Before we can get to reheating, we need to need to specify a Standard Model.  For the purpose of this paper, we will focus on reheating through the coupling to $F\wedge F$.  Because we are building N-flation from the axions associated with four-cycles, this means we should put the standard model on D7 branes.  D3 and D5 branes could be reheated through couplings generated from the superpotential.  But this is likely very difficult to achieve because of the restricted form of the superpotential and we will not consider this possibility in the current discussion.

In order to successfully reheat, the energy from the last few axions (those whose energy density is not significantly diluted by inflation) must be deposited into the Standard Model.  In the above model, this could be a difficult task.  Consider the situation where there are $\tilde{N}+1 \leq N$ independent gauge fields associated with D7 branes wrapping four cycles, with the Standard Model living in one such sector.  The mass basis in which the dynamics of the axions are independent is rotated relative to the basis in which the couplings to gauge fields are simple.  In particular, each gauge field couples to a linear combination of all the different axions of the mass basis.  Therefore, a typical mass basis axion will deposit its energy uniformly into all the different gauge sectors.

Engineering a working model will also be very difficult because of the non-localized nature of the mass basis axions.  In particular, a single axion in this mass basis is related to a four-form potential ($C_4$) that is spread out over the entire compact space.  Therefore, if there are many hidden sectors, there is no obvious way to use locality on the compact space to ensure successful reheating.

Experimentally, the reheating of hidden sectors is constrained only by BBN and over-closure.  For sectors with light particles, BBN requires that the Standard Model dominates the energy density.  For hidden sectors with mass gaps above 1 MeV, the only requirement is that the relic density is, at most, that of the dark matter.  With these constraints, we will determine the tuning of the bases needed for a given hidden sector.  Before we can get to tuning, we need to understand how the relic densities depend on the reheat temperature.  In the following sections, conventions have been chosen to reflect standard references \cite{Kolb:1990vq,Mukhanov:2005sc} as much as possible.

\subsection{Relic Abundances}
Let us take $\tilde{N}$ to be the number of hidden sectors with particles lighter than the reheat temperature.  For sectors with mass gaps above this scale, thermal equilibrium will not be reached and there will be at most some non-thermal production of particles \cite{Chung:1998rq}.  Let $T_{i}$ be the reheat temperature of the $i^{th}$ hidden sector and $T_{SM}$ be the Standard Model reheat temperature.  For our purposes, a hidden sector is a sector which is not in thermal equilibrium with the Standard Model at $T_{SM}$.  Given these definitions, we will not make any assumptions about the precise content of the hidden sectors.  Instead, we will study the constraints on reheating as we vary the general features of their physics.  The hot and cold relic abundances will be estimated using the assumption that $\Gamma = H$ at freeze-out.

In the case of hot relics, the effect is rather simple.  Because the cross section does not enter the relic abundance (except through the constraint that it is relativistic at freeze-out), the abundance is given by the usual formula multiplied by the ratio of energy densities in the two sectors, namely
\beq
\label{hot}
\Omega^{i}_{HR} h^{2}_{75} \simeq \frac{g g'_{FO}}{g_{\star}} \frac{m_{HR}}{19eV} (\frac{T_{i}}{T_{SM}})^3.
\eeq
Here $m_{HR}$ is the mass of the hot relic, $g$ is the number of degrees of freedom of the relic and $g'_{t}= {{g^{SM}_{t} g^{HS}_{RH}}\over {g^{HS}_{t}g^{SM}_{RH}}}$ is a ratio of the effective number of degrees of freedom in the hidden and visible sectors at reheating and at time $t$ (taken in (\ref{hot}) to be the time of freeze-out, $t=FO$).  The effective number of degrees of freedom is defined by $g^{eff}= \sum g_{bos}+ \frac{7}{8}\sum{g_{ferm}}$.  In practice, $g_{t}^{\prime \frac{1}{3}}$ relates the ratios of temperatures at time $t$ to the ratio of temperatures at reheating, which follows from entropy conservation in each sector.  $g_{\star}$ is the effective number of degrees of freedom that convert into Standard Model photons after freeze-out.  For the rest of the paper, we will refer to hot relics as those for which (\ref{hot}) applies.  None of the relics we will discuss are actually `hot' during structure formation or BBN.

Determining when freeze-out occurs for a hot relic is much more model dependant.  Whether or not the particle is relativistic is determined by the hidden sector temperature while expansion is determined by the Standard Model temperature\footnote{Here we assume that the Standard Model dominates the energy density of the universe.  This will be needed later to match BBN predictions.  However, (\ref{hot}) does not depend on this assumption.}.  How the freeze-out temperature scales with the ratio of reheat temperatures will depend upon the dimension of the operator that keeps the relic in thermal equilibrium.

Hot relics will arise in any theory with a mass gap below the reheat temperature.  In such cases, as the universe cools, there are no relativistic species left to carry the energy and the relic abundance of the lightest particle will be given by (\ref{hot}).  Thus, for theories with a mass gap at scales large compared to $1$ MeV this will require substantial tuning of the reheat temperature.

For cold relics, there are competing effects that reduce the dependence on the ratio of temperatures.  In particular, freeze-out is determined by
\beq
n \langle \sigma v \rangle (T_{i}) \simeq H = ({g_{R} \pi^2 \over 90})^{\frac{1}{2}} T_{SM}^2 M_{pl}^{-1}
\eeq
where $g_{R}$ is the number of relativistic species at temperature $T_{SM}$ at freeze-out.  In the above formula, we have assumed that the Standard Model dominates the energy density of the universe. 

Defining $x= {m \over T_{i}}$, then using $\langle \sigma v \rangle \simeq \sigma x^{-\frac{1}{2}}$ and $n \simeq \frac{g}{(2\pi)^{\frac{2}{3}}} T^3 x^{-\frac{2}{3}} e^{-x}$, we can solve for $x$ at freeze-out yielding
\beq
x_{F} \simeq 16.3+ \ln(g g_{R}^{-\frac{1}{2}} ({{g^{\prime \frac{1}{3}} T_{i}} \over T_{SM}})^{2} ({\sigma \over {10^{-38} cm^2}})({m \over {GeV}})).
\eeq
Given the value of $x$ at freeze-out, one can determine the relic abundance.  Noting the extra factor $(\frac{T_{i}}{T_{SM}})^2$, it should not be a surprise that the relic abundance takes the form
\beq
\label{cold}
\Omega^{i}_{CR} h^{2}_{75} \simeq \frac{g g'_{FO}}{g_{\star}} g_{R}^{\frac{1}{2}} (g^{\prime \frac{1}{3}}\frac{T_{i}}{T_{SM}}) x_{F}^{\frac{3}{2}} ({\sigma \over {10^{-38} cm^2}})^{-1}.
\eeq
Because this depends only linearly on temperature, it is not as dramatically affected by changes in the reheat temperature.

\subsection{Aligning Bases}
While aligning the bases of axions with the Standard Model may be extremely difficult to do explicitly in specific cases (just realizing N-flation is difficult in practice \cite{Kallosh:2007ig, KKSS}), we would like to establish how difficult it is in principle.    Here the tuning under discussion is beyond the other types of tuning already required to get the right number of e-folds, etc.  For this purpose, we would like a measure for the alignment of bases.

There is a rather straightforward way to measure the tuning of a given model. Associated with each sector is a single axion $\tilde{\phi}^i$ that couples simply to $F_{i}\wedge F_{i}$.  It can be written as some linear combination of axions in the mass basis as $\tilde{\phi}^{i} = \sum_{j} a^{(i)}_j \phi_{j}$ where the $a^{(i)}_j$ are real coefficients.  We will separate out the overall strength of the coupling (given by $f_{i}$) by requiring $\sum_{j} a^{2 (i)}_j =1$.  This gives a total of $\tilde{N}$ points on $\mathbb{S}^{N-1}$.  However, the relevant question is how the axions whose decay occurs at the end of inflation couple to these sectors.  Axions that decay early have their products diluted by inflation.  Let us denote number of axions whose products contribute to reheating by $M$.  We are only interested in the relative couplings of the hidden sectors to the Standard Model.  An average value of the coupling to the Standard Model should be about $a_{SM}^2=N^{-1}$.  We will absorb this factor into each $f_{i}$ to define the other couplings as points in $\mathbb{B}^M$ (the $M$-dimensional unit ball), assuming that the reheat temperature in the hidden sectors needs to be lower than to the standard model.  This will also imply that the reheat temperature is down by a factor of $N^{-\frac{1}{4}}$.

Constraints on the temperature of hidden sectors can thus be related to constraints on the norm of the vector in $\mathbb{B}^M$.  How tuned the bases must be can then be measured by the available volume after observational constraints.  In order to know how tuned a model is, we need to know how the hidden sector reheat temperature depends on the couplings.  For a partial width of the axion $j$ to the hidden sector $i$, given by $\Gamma^{(i)}_{j} \simeq {{m_{\phi_{j}}^3 a^{2 (i)}_{j}} \over {f_{i}^2}}$, the evolution of the energy density in that sector follows from
\beq
\dot{\rho_{i}}+4 H \rho_{i}=\sum_{j=1}^{M}\Gamma^{(i)}_{j}\rho_{\phi_{j}}.
\eeq
Since $H^2 = \frac{\rho_{tot}}{3 M_{pl}^2}$, when the energy density is dominated by $\rho_{\phi}$ and $\dot{\rho_{i}}$ is small, we can approximate the solution by
\beq
\label{approx}
\rho_{i} \simeq  \sqrt{\frac{3M_{pl}^2}{16\rho_{tot}}}\sum_{j=1}^{M}\Gamma^{(i)}_{j}\rho_{\phi_{j}}.
\eeq
This solution holds until the axions begin to decay dramatically at $t=\Gamma_{tot}^{-1}$.  We will assume the total widths and energy densities for these axions are approximately equal.  Using $\rho_{i} \propto T^4$ and $\rho_{\phi} \propto t^{-2}$, the solution (\ref{approx}) at $t=\Gamma_{tot}^{-1}$ yields the reheat temperature
\beq
\label{ti}
T_{i} \propto ({{\Gamma^{(i)} \Gamma_{tot}} \over {r_{i}}})^{\frac{1}{4}},
\eeq
where $\Gamma^{(i)}$ is the sum of the widths from the remaining $M$ axions and $r_{i}$ is the number of relativistic degrees of freedom in sector $i$ at reheating.  In the case where there are no hidden sectors, and thus $\Gamma^{(i)} =  \Gamma_{tot}$, this reproduces the usual formula.  As a result, $r_{i} T^4_{i}\propto \sum_{j=1}^{M} a^{2 (i)}_{j}$.

Let us now return to the question of tuning and generality.  First consider the case where all the hidden sectors have massless particles and no over-closure problems.  The only constraint is that the energy density during BBN is dominated by relativistic Standard Model particles \cite{Fields:2006ga,Steigman:2005uz,Sarkar:1995dd}.  Therefore, we will take a modest estimate that $r_{SM}(T_{SM}^4)_{BBN} \geq (10^{2} \tilde{N} r_{i} T^{4}_{i})_{BBN}$, for all $i$.  Because the temperature of radiation in any sector during BBN also depends on the number of species that decay to radiation before $T \approx 1$ MeV, we must include a factor of $g_{BBN}^{\prime \frac{1}{3}}$ in each sector when comparing the reheat temperatures.  Therefore, for each hidden sector, the constraint on $a^{(i)}_{j}$ is that it lies inside a ball such that $R^2 \leq 10^{-2} \tilde{N}^{-1} r_{i}^{\prime}g^{\prime -\frac{4}{3}}_{i}$, where $r_{i}^{\prime} = (\frac{r_{SM}}{r_{i}})_{BBN}(\frac{r_{i}}{r_{SM}})_{RH}$.  There is a minor constraint on the standard model vector, but we will ignore this contribution as it has little effect.  Defining the tuning $\zeta$ to be the ratio of available volume to the total, the tuning is
\beq
\label{tune1}
\zeta \simeq (10^{-2} \tilde{N}^{-1})^{\frac{\tilde{N}M}{2}} \prod_{i=1}^{\tilde{N}} r_{i}^{\prime \frac{M}{2}}(g^{\prime -\frac{2M}{3}}_{i})_{BBN}.
\eeq
Note that neither $M$ nor $\tilde{N}$ are necessarily on the order of $N$, which must be large to get 60 efolds.  $M$ is the number of axions relevant to the very final stages and is likely between one and five (knowing this likely requires simulations, but we will remain agnostic about the precise value).  $\tilde{N}$ is the number of hidden sectors with mass below the scale of reheating.  It is unknown what is generic for $\tilde{N}$.  One can arrange for $\tilde{N}=0$, but it can also be on the order of $N$.

To get a feel for the tuning formulas, let us determine the tuning in a couple of simple cases.  First of all, it should be pointed out that the two ratios of degrees of freedom, $r'$ and $g'$ are competing effects, which renders them relatively insignificant.  This can be seen as follows:  by increasing the effective number of degrees of freedom at reheating that decay to the Standard Model, the energy density at BBN increases by $g'^{4 \over 3}$.  However, the increase in relativistic number of degrees of freedom lowers $T^{4}_{SM}$ by $r'^{-1}$.  Thus, even if $r'\simeq g' \simeq 10^{-3}$ for every hidden sector, the net effect on $\zeta$ is $\sqrt{10}^{M \tilde{N}}$.  At best, this contribution adjusts the numerical value $10^{-2}$ that already appears in (\ref{tune1}).  It should be clear the dominant effect comes from $\tilde{N}$.  For a single hidden sector ($\tilde{N}=1$) the tune isn't particularly bad, as $\zeta \simeq 10^{-1}$ to $\zeta \simeq 10^{-5}$ for $M=1 - 5$ and $r'=g'=1$.  $\zeta$ can be order one if $r'\simeq g' \simeq 10^{-3}$.  However, even at $M=1$, if $\tilde{N}=10-100$ then $\zeta \simeq 10^{-15}-10^{-200}$.  Thus, using BBN, the tuning is only severe if we have a very large number of hidden sectors.

Avoiding over-closure will require further tuning when the hidden sectors leave large relic abundances.  Hot relics are the most likely concern, as any hidden sector with a mass gap below the reheat temperature will be dangerous.  Because (\ref{hot}) depends on the ratio of temperatures cubed, it will be easier in most cases to lower the relic abundance rather than to avoid initial production.  For hot relics with masses greater than $1$ MeV, this constraint is more severe than BBN.  Given $P \leq \tilde{N}$ sectors with hot relics of mass $m_{i}$, then the tuning becomes
\beq
\label{tune2}
\zeta \simeq (10^{-2} \tilde{N}^{-1})^{\frac{(\tilde{N}-P)M}{2}}\prod_{j=P+1}^{\tilde{N}}r_{j}^{\prime \frac{M}{2}}(g^{\prime -\frac{2M}{3}}_{j})_{BBN}\prod_{i=1}^{P}(\frac{g_{\star}}{g g'}\frac{19 eV}{m_i})^{\frac{2M}{3}}.
\eeq
The appearance of 19 eV should not be taken too seriously, given that there is significant dependence on degrees of freedom in both the visible and hidden sectors.  For masses greater than $10^3$ TeV, it may be easier to lower the temperature below the mass scale of the relic particle altogether.   Then we would have
\beq
\label{tune3}
\zeta \simeq (10^{-2}\tilde{N}^{-1})^{\frac{(\tilde{N}-P)M}{2}}\prod_{j=P+1}^{\tilde{N}}r_{j}^{\prime \frac{M}{2}}(g^{\prime -\frac{2M}{3}}_{j})_{BBN}\prod_{i=1}^{P}(\frac{r_{i}}{r_{SM}})^{\frac{M}{2}}_{RH}(\frac{m_i}{T_{SM}})^{2M}.
\eeq
Given a typical reheat temperature for N-flation of $T_{SM} \simeq 10^{10}$ GeV, the worst tuning arises at the crossover from (\ref{tune2}) to (\ref{tune3}) when $m = 10^3$ TeV.  Thus, the cost of tuning for these relics is no worse than $10^{-8}$ for each hidden sector.

What is generic ultimately depends on what constitutes generic particle content for a hidden sector.  If theories with mass gaps below $10^3$ TeV are generic, then so too are warm hidden sectors.  For such models, hidden sector dark matter also becomes a possibility, as the dark matter from hot relics with mass greater than 1 MeV is consistent with data from both the CMB and structure formation (in fact, small scale structure may favor warm dark matter \cite{Colombi:1995ze, Bode:2000gq}).

\section{Generalizations}
Thus far, very little of what we have done is specific to N-flation.  In particular, in a model which has a single inflaton with axion-like couplings, the above formulas apply with $M=1$.  In known closed string models 
\cite{Blanco-Pillado:2004ns, Conlon:2005jm, Bond:2006nc,Blanco-Pillado:2006he} the above discussion typically applies.  However, the problem seems more general than the particular form of coupling, as it is simply an issue of localizing the decay of the inflaton.

If we focus on type II models with closed strings acting as the inflaton, it appears that the same issues that occur in reheating N-flation persist.  In any such constructions, the Standard Model will live on D-branes which are localized in the compact space; the inflaton, however, is spread out over the entire space.  Successful reheating will ultimately require one to localize the inflaton decay onto the Standard Model brane, rather than one of the hidden sectors.  While it is likely the case that one can engineer models in which this occurs (this could be done most easily by massing up all the hidden sectors), it is not clear that such models are generic, or that they survive other constraints coming from self consistency, cosmology, and particle physics.

The tuning in more general closed string models can also be studied.  Whenever the decay of the inflaton is perturbative, the tuning formulas still apply for any sectors that reheat through operators of the same dimension as the coupling to the Standard Model.  The only difference in those cases will be the reheat temperature. This will change the scale where the cross-over between (\ref{tune2}) and (\ref{tune3}) occurs.  In cases where the reheating proceeds non-perturbatively (e.g. \cite{Traschen:1990sw,Kofman:1994rk,Felder:1998vq,Greene:1998nh,Baacke:1998di}), the general problem remains, but the same formulas no longer apply.  In particular, the reheat temperatures have a very nontrivial dependence on the couplings which makes tunings difficult to calculate explicitly.

These considerations can become particularly important when considering SUSY model building.  Such models typically require SUSY-breaking sectors, which may contain stable particles.  These sectors play an important role in phenomenology as they (along with the mediation mechanism) determine the soft SUSY-breaking parameters in the MSSM.   If these sectors are reheated and do not decay before BBN then for such models, we will need to tune the hidden sector reheat temperature.  This is in addition to Standard Model reheat temperature which is usually bounded by overproduction of gravitinos, etc.  In general, one can state the problem as being that the inflaton need not couple only to the Standard Model.

One could also take the view that the above arguments suggest that open string inflation (e.g. \cite{Baumann:2007ah,Kachru:2003sx,Alishahiha:2004eh,Silverstein:2003hf}) is a more natural candidate.  In some such cases, reheating is accomplished through D-brane annihilation and has been well studied.  When the Standard Model throat and inflationary throat are the same, it is clear how this model is localized to reheat the Standard Model preferentially.  However, experimental signatures such as relic KK modes \cite{Chen:2006ni} and cosmic strings \cite{Sarangi:2002yt,Copeland:2003bj,Dvali:2003zj} could distinguish them from the more garden-variety reheating of closed string models.  Additionally, there is the possibility of producing gravitons at reheating that might be observable in the future.  For this reason, it would be worthwhile to determine the precise gravitational wave signature of these models.  

\section{Conclusion}
Motivated by the difficulties encountered in building natural models of inflation in UV-complete theories, we have studied reheating in closed string models of inflation.  In type II models, the Standard Model and hidden sectors live on D-branes, which are localized in the compact space.  Because the inflaton is generally not localized (not even approximately), the hidden sectors are reheated along with the Standard Model.  Bounds from BBN and over-closure can be avoided when the inflaton is tuned to decay preferentially onto the Standard Model branes.  For reheating through elementary decays, the amount of tuning necessary was calculated.  Similar tuning requirements will arise when reheating is non-perturbative, although the tunings in that case are difficult to calculate.  In all cases, the tuning depends on general features of the hidden sector, such as the size of the mass gap and the number of independent sectors.

To the author's knowledge, there is no consensus on what is generic for the hidden sectors of realistic compactifications.  The tuning is only severe in cases with large numbers of independent sectors and/or cases where sectors have masses well below the reheat temperature but well above 1 MeV.  For sectors with gaps below $10^3$ TeV (when $T_{SM} \simeq 10^{10}$ GeV), observational bounds are most easily met by allowing those sectors to reheat and having the lowest mass particle be the dark matter.  When the gap is above this scale, the sector must not be reheated.

If we take the view that current string models tell us where to look for the effects of UV complete theories, then reheating string models points to hidden sector matter as a possibly generic outcome.  As data on both dark matter and gravitational waves improves in the coming years, we may hope to see signatures of this UV physics.

\acknowledgments
I would to thank Shamit Kachru, Liam McAllister, Eva Silverstein, Peter Svrcek, and Jay Wacker for helpful discussions.  This project was supported in part by a NSERC Postgraduate fellowship, by the DOE under contract DE-AC03-76SF00515 and by the NSF under contract 9870115.

\bibliography{reheatbib}

\end{document}